\documentclass[prd,superscriptaddress,preprint,aps,amsmath,amssymb,showpacs,nofootinbib]{revtex4-2}

\usepackage{graphicx,enumerate,xcolor,enumitem,xspace,braket}
\usepackage[colorlinks=true,citecolor=blue]{hyperref}

\usepackage{comment}
\usepackage{multirow}
\usepackage{subfig}
\usepackage{tikz-feynman}
\usepackage{caption}
\usepackage[normalem]{ulem}

\begin{document}

\title{Entanglement and Bell Nonlocality in $\tau^+ \tau^-$ at the BEPC}

\author{Tao Han}
\email{than@pitt.edu}
\affiliation{PITT PACC, Department of Physics and Astronomy,\\ University of Pittsburgh, 3941 O’Hara St., Pittsburgh, PA 15260, USA}

\author{Matthew Low}
\email{matthew.w.low@pitt.edu}
\affiliation{PITT PACC, Department of Physics and Astronomy,\\ University of Pittsburgh, 3941 O’Hara St., Pittsburgh, PA 15260, USA}

\author{Youle Su}
\email{ylsu21@m.fudan.edu.cn}
\affiliation{Department of Physics and Center for Field Theory and Particle Physics, \\ Fudan University, Shanghai 200438, China} 

\date{\today}
{\hfill PITT-PACC-2412}

\begin{abstract}
Quantum entanglement and Bell nonlocality are two phenomena that occur only in quantum systems.  In both cases, these are correlations between two subsystems that are classically absent.  Traditionally, these phenomena have been measured in low-energy photon and electron experiments, but more recently they have also been measured in high-energy particle collider environments.  In this work, we propose measuring the entanglement and Bell nonlocality in the $\tau^+\tau^-$ state near and above its kinematic threshold at the Beijing Electron Positron Collider (BEPC).   We find that in the existing dataset, entanglement is observable if systematic uncertainties are kept to $1\%$.  In the upcoming run between 4.0 and 5.6 GeV, the entanglement is predicted to be measurable with a precision better than 4\% and Bell nonlocality can be established at $5\sigma$ as long as systematic uncertainty can be controlled at level of $0.5\% -2.0\%$, depending on the center-of-mass energy.
\end{abstract}

\maketitle
\newpage
{
\hypersetup{linkcolor=blue}
\tableofcontents
}

\section{Introduction}
\label{sec:intro}

Quantum mechanics and special relativity form the cornerstones of modern physics.  Until recently, the measurement and manipulation of quantum mechanical systems have been performed by specifically designed experiments and primarily in low-energy systems.  In 2020, it was shown that at the CERN Large Hadron Collider (LHC), the multipurpose detectors, designed to study high-energy particles, could explicitly measure the entanglement between the spin of a top quark and the spin of an anti-top quark~\cite{Afik:2020onf}.  Other work followed that showed that in addition to entanglement, Bell nonlocality could be measured in the $t\bar{t}$ system~\cite{Fabbrichesi:2021npl,Severi:2021cnj,Aoude:2022imd,Afik:2022kwm,Aguilar-Saavedra:2022uye,Fabbrichesi:2022ovb,Severi:2022qjy,Dong:2023xiw,Han:2023fci}. In fact, the sizable production rate of $t\bar{t}$, its relatively large mass, its rapid decay, and its simple quantum-mechanical description make it an ideal system for quantum information in the LHC.  Measurements of entanglement in this system have already been made by both ATLAS~\cite{ATLAS:2023fsd} and CMS~\cite{CMS:2024pts}. Interest has been growing to explore other quantum informational quantities in $t\bar{t}$~\cite{Afik:2022dgh,Aguilar-Saavedra:2023hss,Aguilar-Saavedra:2024fig,White:2024nuc,Han:2024ugl,Fabbrichesi:2025ywl} and other final states~\cite{Ashby-Pickering:2022umy,Aguilar-Saavedra:2022wam,Fabbrichesi:2023cev,Aguilar-Saavedra:2022mpg,Altakach:2022ywa,Aoude:2023hxv,Morales:2023gow,Barr:2024djo,Bernal:2023ruk,Bi:2023uop,Ma:2023yvd,Sakurai:2023nsc,Bernal:2023jba,Han:2023fci,Ehataht:2023zzt,Cheng:2023qmz,Aguilar-Saavedra:2024hwd,Barr:2024djo,Aguilar-Saavedra:2024vpd,Aguilar-Saavedra:2024whi,Duch:2024pwm,Morales:2024jhj,Subba:2024mnl,Maltoni:2024csn,Afik:2024uif,Wu:2024asu,Cheng:2024btk,Gabrielli:2024kbz,Ruzi:2024cbt,Cheng:2024rxi,Wu:2024ovc,Gao:2024leu,Altomonte:2024upf,Fabbrichesi:2024rec,Cheng:2025cuv} at colliders.

In parallel with this exploration work, there have been important advances in optimizing the methods for extracting quantum information from collider environments.  One of these was the identification of fictitious states~\cite{Afik:2022kwm,Cheng:2023qmz,Cheng:2024btk} which are ubiquitous in colliders.  Ideally, one would reconstruct the underlying quantum state at colliders; however, because of averaging, one actually reconstructs a fictitious state.  These are still useful in most contexts and still demonstrate the existence of entanglement and Bell nonlocality. An advantageous outcome of the use of fictitious states is that one can select a spin quantization basis that optimizes the signal size of the entanglement or Bell nonlocality~\cite{Cheng:2023qmz,Cheng:2024btk}.

A second important advance is the development of the kinematic method in reconstructing the density matrix~\cite{Cheng:2024rxi}.  The majority of past work utilizes the decay method which requires the qubit particles to decay.  The angles of decay products are then used to reconstruct the quantum density matrix because the decay product angles are correlated with the qubit spins.  This method requires a reasonable assumption of the spin properties of the decaying particles.  The kinematic method, in contrast, uses the reasonable assumption of the production mechanism of qubit particles.  This allows the quantum density matrix to be reconstructed even when the qubit particles do not decay.  In this work, we will compare both methods and show that the kinematic method provides drastically smaller statistical uncertainties.  For the system we study, the use of the kinematic method is critical to achieving discovery-level results.

With well-constrained kinematics and beam control of the initial states, $e^+e^-$ colliders offer another environment to study quantum information. There were early attempts to test Bell nonlocality at LEP experiments \cite{Privitera:1991nz,Abel:1992kz,Dreiner:1992gt}.
Recent work has shown that Bell nonlocality is potentially observable in flavor oscillations in Belle II~\cite{Chen:2024drt}.   More relevantly for this work, it was demonstrated that the bipartite qubit system of $\tau^+ \tau^-$ exhibits both entanglement and Bell nonlocality over a wide range of energies in the future $e^+e^-$ colliders~\cite{Fabbrichesi:2024wcd}.  Other work on the $\tau^- \tau^+$ final state includes Refs.~\cite{Fabbrichesi:2022ovb,Ma:2023yvd,Ehataht:2023zzt,LoChiatto:2024dmx} (see also Ref.~\cite{Banerjee:2023qjc}).

In this paper, we set out to explore the sensitivity to measure quantum entanglement and Bell nonlocality with the process $e^+ e^- \to \tau^+ \tau^-$ in the Beijing Spectrometer experiment (BES-III) at the Beijing Electron-Positron Collider (BEPC-II).  During the past three decades, the BES experiments at the BEPC have made great achievements in $\tau$-charm physics~\cite{BESIII:2020nme,BESIII:2022mxl}.  The BES-III experiment has made record-setting measurements of the $\tau$ mass~\cite{BESIII:2014srs} and the QCD $R$ value~\cite{BES:2001ckj}, collected an enormous sample of $\psi(2S)$ mesons~\cite{BESIII:2024lks}, delivered rich charm physics \cite{BESIII:2022mxl}, and made important observations of exotic hadronic states~\cite{BES:2005ega,BESIII:2010gmv}.  The center-of-mass energies near and above the $\tau^+ \tau^-$ threshold and the anticipated large data sample at energies between 4.0 and 5.6 GeV in the BES-III experiment strongly motivate the exploration of quantum information in such a mature research program. Past work identifying BEPC-II as a promising environment for quantum information includes a study on hyperon decays~\cite{Wu:2024asu} and charmed baryon decays~\cite{Fabbrichesi:2024rec}.

Exploring the large data sample of $\tau^+\tau^-$ in BES-III, we find that the reach for entanglement ranges from a $1.5\sigma$ detection with a $5\%$ systematic uncertainty to a $7\%$ precision with a $0.5\%$ systematic uncertainty.  This requires the use of the kinematic method mentioned above.  The current dataset does not have sensitivity to Bell nonlocality.  In an upcoming dataset at energies between 4.6 and 5.0 GeV, entanglement should be detected at $>5\sigma$ even with systematic uncertainties as large as $5\%$.  Bell nonlocality, on the other hand, should be observable at $>5\sigma$ provided the systematic uncertainties are less than $2\%$ and the use of the kinematic method.

The remainder of the paper is organized as follows. We first outline the general features of quantum information in the process $e^+ e^- \to \tau^+ \tau^-$ in Sec.~\ref{sec:tautau}.  We present details for $\tau$ reconstruction and a proposed analysis in the BES-III experiments with the current and anticipated energy plans at BEPC-II in Sec.~\ref{sec:BEPC}.  We summarize our results and conclude in Sec.~\ref{sec:summary}.  Two appendices provide additional details.

\section{Quantum Tomography for $e^+ e^- \to \tau^+ \tau^-$}
\label{sec:tautau}

\subsection{Quantum Information}
\label{sec:qi}

In the Standard Model (SM) of particle physics, the process of $e^+ e^- \to \tau^+ \tau^-$ proceeds with two diagrams, one with an $s$-channel photon and one with an $s$-channel $Z$ boson.  The $\tau$ lepton is a spin-1/2 particle, corresponding to a qubit, such that the $\tau^+ \tau^-$ can form a bipartite qubit system.  For simplicity, we show most of our analytical expressions only with the leading-photon contribution.  However, for all numerical results, the $Z$ boson is included.

As a quantum state, this $\tau^+ \tau^-$ system is fully described by a $4 \times 4$ density matrix $\rho$.  A useful parametrization of this system is the Fano-Bloch decomposition~\cite{Fano:1983zz}
\begin{equation} \label{eq:fano}
\rho = \frac{1}{4}\left(
\mathbb{I}_2 \otimes \mathbb{I}_2
+ \sum_i B^+_i \sigma_i \otimes \mathbb{I}_2
+ \sum_j B^-_j \mathbb{I}_2 \otimes \sigma_i
+ \sum_{ij} C_{ij} \sigma_i \otimes \sigma_j
\right),
\end{equation}
where the sums run over $i,j = 1,2,3$.  Above, $\mathbb{I}_2$ is the two-dimensional identity matrix and $\sigma_i$ are the Pauli matrices.  The coefficient $B^+_i$ is the net polarization of the first qubit, $B^-_j$ is the net polarization of the second qubit, and $C_{ij}$ is the spin correlation matrix.  Invariance under CP enforces $B^+ = B^-$ and $C = C^T$.  Since the $\tau^+ \tau^-$ system is produced from the electroweak interaction, the polarization in general is non-zero.

As can be seen by the appearance of Pauli matrices in Eq.~\eqref{eq:fano}, it is necessary to choose a basis with which to quantize the spin.  The beam basis $\{ \hat{x}, \hat{y}, \hat{z} \}$ is a common choice where $\hat{z}$ points along the $e^-$ momentum direction and $\hat{x}$ and $\hat{y}$ span the transverse plane. Another common choice is the helicity basis $\{ \hat{k}, \hat{r}, \hat{n} \}$ where $\hat{k}$ points along the direction of motion of the $\tau^-$, $\hat{r} = (\hat{z} - \hat{k} \cos\theta) / \sin\theta$, and $\hat{n} = \hat{r} \times \hat{k}$.  In this work, we use the diagonal basis~\cite{Cheng:2023qmz,Cheng:2024btk} which is defined as the basis that makes the spin correlations maximal.  We will define this basis in Sec.~\ref{sec:analysis}.

In the diagonal basis, there is a useful low-energy parameterization of the density matrix.  We first define the mixed state $\rho_{\rm mixed}$ and pure state $\rho_{\rm pure}$ by
\begin{align}
\label{eq:rhos}
\rho_{\rm mixed} &= \frac{1}{2} \ket{\uparrow \uparrow} \bra{\uparrow \uparrow}  
+ \frac{1}{2} \ket{\downarrow \downarrow} \bra{\downarrow \downarrow}, \\ 
\rho_{\rm pure} &= \ket{\psi^+} \bra{\psi^+}, \qquad  \ket{\psi^+} = \frac{1}{\sqrt{2}}( \ket{\uparrow\uparrow} + \ket{\downarrow\downarrow} ),
\end{align}
where $\uparrow$ ($\downarrow$) is the $+$ ($-$) eigenvalue of the $\hat{e}_3$ direction in the diagonal basis.  The general density matrix of the system is then
\begin{equation}
\label{eq:rho_decomp}
\rho = (1 - \lambda) \rho_{\rm mixed}
+ \lambda \rho_{\rm pure},
\qquad
\lambda = \frac{\beta^2 \sin^2 \theta}{2 - \beta^2 \sin^2 \theta},
\end{equation}
where $\beta$ is the velocity in the center-of-mass frame of the $\tau$, and $\theta$ is the scattering angle in the center-of-mass frame, with respect to the beam.  Beyond the existing result for $\theta = \pi/2$ in Ref.~\cite{Fabbrichesi:2022ovb}, we extend Eq.~\eqref{eq:rho_decomp} over the full phase space.

A quantum state $\rho$ is separable when it has the following decomposition.
\begin{equation}
\rho = \sum_i p_i \rho_{A,i} \otimes \rho_{B,i},
\qquad\qquad
\sum_i p_i = 1,
\end{equation}
where $\rho_{A,i}$ is a $2 \times 2$ matrix that describes subsystem $A$ and $\rho_{B,i}$ is a $2 \times 2$ matrix that describes subsystem $B$.  If a state is not separable, it is entangled. Entangled states exhibit a degree of correlation that is not attainable in classical systems~\cite{Einstein:1935rr}.

One measure of entanglement is the concurrence, which for a bipartite qubit system is~\cite{Wootters:1997id}
\begin{equation} 
\label{eq:concurrence}
\mathcal{C}(\rho) = \text{max}(0,\lambda_1 - \lambda_2 - \lambda_3 - \lambda_4),
\end{equation}
where $\lambda_i$ are the eigenvalues, ordered from largest to smallest, of the auxiliary matrix $R$ 
\begin{equation}
R = \sqrt{\sqrt{\rho} \tilde{\rho} \sqrt{\rho}},
\qquad\qquad
\tilde{\rho} = (\sigma_2 \otimes \sigma_2) \rho^* (\sigma_2 \otimes \sigma_2).
\end{equation}
Separable states have $\mathcal{C}(\rho)=0$, while entangled states have $0 < \mathcal{C}(\rho) \leq 1$.  A higher value of $\mathcal{C}(\rho)$ indicates a more entangled state.

At center-of-mass energies, $\sqrt{s}$, relevant for BEPC-II, the polarization components $B^+_i$ and $B^-_j$ scale as $s/m_Z^2 \sim 10^{-4}$ so we can safely neglect these and use the simplified formula  
\begin{equation}
\label{eq:concurrence_linear}
\mathcal{C}(\rho) = \frac{1}{2}\left (C_{11} + C_{33}-C_{22}-1\right).
\end{equation}
Bell's inequality is an equation that distinguishes Bell local quantum states from Bell nonlocal quantum states.  Bell nonlocal states exhibit a particular strong quantum correlation that enables quantum teleportion~\cite{Bennett:1992tv} among many other phenemona.

For a bipartite qubit system, Bell's inequality is the Clauser-Horne-Shimony-Holt (CHSH) inequality~\cite{Clauser:1969ny}
\begin{equation} \label{eq:CHSH}
| \langle \vec{a}_1 \cdot \vec{\sigma} \otimes \vec{b}_1 \cdot \vec{\sigma} \rangle
- \langle \vec{a}_1 \cdot \vec{\sigma} \otimes \vec{b}_2 \cdot \vec{\sigma} \rangle
+ \langle \vec{a}_2 \cdot \vec{\sigma} \otimes \vec{b}_1 \cdot \vec{\sigma} \rangle
+\langle \vec{a}_2 \cdot \vec{\sigma} \otimes \vec{b}_2 \cdot \vec{\sigma} \rangle |
\leq 2.
\end{equation}
The three-vectors $\vec{a}_1$, $\vec{a}_2$, $\vec{b}_1$, and $\vec{b}_2$ are the measurement axes or ``detector settings.''  In a traditional 
low-energy experiment the first term of Eq.~\eqref{eq:CHSH} corresponds to running the experiment with the detector for the first qubit with setting $\vec{a}_1$ and the detector for the second qubit with setting $\vec{b}_1$.  Data is then taken separately for each term with the appropriate detector settings.

In the collider environment, each term corresponds to a spin measurement with the quantization axis specified by the appropriate vectors: $\vec{a}_1$, $\vec{a}_2$, $\vec{b}_1$, and $\vec{b}_2$.  Each term of Eq.~\eqref{eq:CHSH} is a simultaneous spin measurement of each qubit which corresponds to an entry (or linear combination of entries) of the spin correlation matrix.

We label the left-hand side of Eq.~\eqref{eq:CHSH} as the Bell variable $\mathcal{B}$
\begin{equation}
\mathcal{B}(\vec{a}_1, \vec{a}_2, \vec{b}_1,\vec{b}_2) = |\langle \vec{a}_1 \cdot \vec{\sigma} \otimes \vec{b}_1 \cdot \vec{\sigma} \rangle
- \langle \vec{a}_1 \cdot \vec{\sigma} \otimes \vec{b}_2 \cdot \vec{\sigma} \rangle
+ \langle \vec{a}_2 \cdot \vec{\sigma} \otimes \vec{b}_1 \cdot \vec{\sigma} \rangle
+\langle \vec{a}_2 \cdot \vec{\sigma} \otimes \vec{b}_2 \cdot \vec{\sigma} \rangle|.
\end{equation}
The CHSH inequality becomes $\mathcal{B}(\vec{a}_1, \vec{a}_2, \vec{b}_1,\vec{b}_2) < 2$.  In order to identify a quantum state as Bell nonlocal, $\vec{a}_1$, $\vec{a}_2$, $\vec{b}_1$, and $\vec{b}_2$ should be chosen to maximize $\mathcal{B}$.  The optimal value is given by~\cite{Horodecki:1995nsk,Fabbrichesi:2021npl}
\begin{equation}
\label{eq:bellnonlocality}
\mathcal{B}_{\rm max} = 2 \sqrt{m_1 + m_2},
\end{equation}
where $m_1$ and $m_2$ are the largest and second largest eigenvalues, respectively, of $C^T C$.  The violation of Bell's inequality corresponds to $\mathcal{B}_{\rm max} > 2$.

It is often simpler to work with a linear approximation of $\mathcal{B}_{\rm max}$ that is~\cite{Severi:2021cnj}
\begin{equation}
\mathcal{B}_{\rm lin} = \text{max}_{ij} \sqrt{2}\ | C_{ii} \pm C_{jj}|,
\qquad\qquad
(i,j=1,2,3).
\end{equation}
In this work, we use $\mathcal{B} = \mathcal{B}_{\rm max}$.

From Eq.~\eqref{eq:rho_decomp}, we can gain some intuitive understanding of the expected results of the system.  At threshold $\beta \to 0$, the $\tau^+ \tau^-$ state consists mainly of $\rho_{\rm mixed}$ which is separable, thus the $\tau^+ \tau^-$ state is neither entangled nor Bell nonlocal.  As $\beta \to 1$, the $\tau^+ \tau^-$ state approaches $\rho_{\rm pure}$, which is a Bell state and consequently has maximal entanglement and maximal Bell nonlocality.   The energies of BEPC-II do not extend far above the $\tau^+ \tau^-$ threshold, which leads to a small signal both of concurrence and of Bell nonlocality.  Their detectability depends on the uncertainties of the measurements, which will be calculated in Sec.~\ref{sec:BEPC}.

\subsection{$\tau^+\tau^-$ Production in $e^+e^-$ Collisions}
\label{sec:process}

The spin correlation matrix for the unpolarized process $e^+ e^- \to \tau^+ \tau^-$ is given by~\cite{Maltoni:2024csn}
\begin{subequations}
\label{eq:fano_helicity}
\begin{align}
C_{11} &= \frac{1}{c_0} (F^{[0]} \left (\beta^2-(\beta^2-2)\cos^2\theta\right )+2F^{[1]}\cos\theta+F^{[2]}(1+\cos^2\theta)), \\
C_{13} &= \frac{1}{c_0}(-2 F^{[0]} \sqrt{1-\beta^2}\sin\theta\cos\theta-F^{[1]}\sqrt{1-\beta^2}\sin\theta), \\
C_{22} &= \frac{1}{c_0}(-F^{[0]}\beta^2\sin^2\theta+F^{[2]}\sin^2\theta), \\
C_{33} &= \frac{1}{c_0} (F^{[0]}(2-\beta^2)\sin^2\theta-F^{[2]}\sin^2\theta).
\end{align}
\end{subequations}
Furthermore, $C_{31} = C_{13}$ and $C_{12} = C_{21} = C_{23} = C_{32} = 0$.  The other factors are
\begin{align}
c_0 &= F^{[0]}(\beta^2\cos^2\theta-\beta^2+2)+2F^{[1]}\cos\theta+F^{[2]}(1+\cos^2\theta), \\
F^{[0]} &= 48\left(Q_{\mathrm{\tau}}^2 Q_{e}^2+2 \operatorname{Re} \frac{4 Q_{\mathrm{\tau}} Q_{e} g_{\mathrm{V\tau}} g_{\mathrm{V} e} m_{\mathrm{\tau}}^2}{c_{\mathrm{W}}^2 s_{\mathrm{W}}^2\left(4 m_{\mathrm{\tau}}^2-\left(1-\beta^2\right) m_{\mathrm{Z}}^2\right)}+\frac{16 g_{\mathrm{V\tau}}^2 m_{\mathrm{\tau}}^4\left(g_{\mathrm{V} e}^2+g_{\mathrm{A} e}^2\right)}{\left|c_{\mathrm{W}}^2 s_{\mathrm{W}}^2\left(4 m_{\mathrm{\tau}}^2-\left(1-\beta^2\right) m_{\mathrm{Z}}^2\right)\right|^2}\right), \\
F^{[1]} &= 192 g_{\mathrm{A\tau}} g_{\mathrm{A} e} m_{\mathrm{\tau}}^2 \beta\left(\frac{16 g_{\mathrm{V\tau}} g_{\mathrm{V} e} m_{\mathrm{\tau}}^2}{\left|c_{\mathrm{W}}^2 s_{\mathrm{W}}^2\left(4 m_{\mathrm{\tau}}^2-\left(1-\beta^2\right) m_{\mathrm{Z}}^2\right)\right|^2}+2 \operatorname{Re} \frac{Q_{\mathrm{\tau}} Q_{e}}{c_{\mathrm{W}}^2 s_{\mathrm{W}}^2\left(4 m_{\mathrm{\tau}}^2-\left(1-\beta^2\right) m_{\mathrm{Z}}^2\right)}\right), \\ 
F^{[2]} &= \frac{768 g_{\mathrm{A\tau}}^2 m_{\mathrm{\tau}}^4 \beta^2\left(g_{\mathrm{V} e}^2+g_{\mathrm{A} e}^2\right)}{\left|c_{\mathrm{W}}^2 s_{\mathrm{W}}^2\left(4 m_{\mathrm{\tau}}^2-\left(1-\beta^2\right) m_{\mathrm{Z}}^2\right)\right|^2},
\end{align}
where $s_W$ and $c_W$ are the sine and cosine of the weak mixing angle, respectively.  The couplings are $Q_\tau = Q_e = -1$, $g_{V\tau} = g_{Ve} = I_3/2 - Q_e s_W^2 = -0.019$, and $g_{A\tau} = g_{Ae} = I_3/2 = - 0.25$. 

At low energies $\sqrt{s} \ll m_Z$, the spin correlation matrix simplifies to~\cite{Ehataht:2023zzt}
\begin{equation}
\label{eq:spincorrelationmatrix}
    C_{ij}=\frac{1}{2-\beta^2\sin^2\theta}\left (\begin{array}{ccc}
        \left (2-\beta^2\right )\sin^2\theta & 0 & \sqrt{1-\beta^2}\sin\left (2\theta\right ) \\
        0 & -\beta^2\sin^2\theta & 0 \\
        \sqrt{1-\beta^2}\sin\left (2\theta\right ) & 0 & \beta^2+\left (2-\beta^2\right )\cos^2\theta
    \end{array}\right )_{ij}.
\end{equation}
This leads to the concurrence $\mathcal{C}(\rho)$ in Eq.~(\ref{eq:concurrence_linear}) as 
\begin{equation}
\label{eq:kin_c}
\mathcal{C}(\rho)=\frac{\beta^2\sin^2\theta}{2-\beta^2\sin^2\theta}.
\end{equation}
The Bell variable $\mathcal{B}(\rho)$ is
\begin{equation}
\label{eq:kin_b}
\mathcal{B}(\rho)=2\sqrt{1+\left (\frac{\beta^2\sin^2\theta}{2-\beta^2\sin^2\theta}\right )^2}\ . 
\end{equation}
One may observe there is a simple relation between the mixing parameter $\lambda$ from Eq.~\eqref{eq:rho_decomp} and the concurrence, $\mathcal{C}(\rho)=\lambda$, and Bell variable, $\mathcal{B}(\rho)= 2\sqrt{1+\lambda^2}$.  For states that are parametrized in a simple way by forms similar to Eq.~\eqref{eq:rho_decomp}, there often exist simple relations between various quantum properties.

\subsection{Decays of the $\tau$}
\label{sec:decays}

To obtain the density matrix of the $\tau^+\tau^-$ state, it is necessary to reconstruct $\tau$.  For the kinematic method, we only need to know the speed $\beta$ and the scattering polar angle $\theta$ of the $\tau$. For the decay method, we additionally need to measure the decay angle of one of the decay products in the $\tau$ rest frame, which is a proxy for the spin of the $\tau$.

Consider the differential decay of $\tau$ in its rest frame~\cite{Tsai:1971vv,Kuhn:1995nn,Bullock:1992yt,L3:1998oan}
\begin{equation}
\frac{1}{\Gamma} \frac{d\Gamma}{d\cos\theta_d}
= \frac{1}{2} \left( 1 + P \kappa \cos\theta_d \right).
\label{eq:spin}
\end{equation}
The angle $\theta_d$ is between the selected decay product and the polarization axis of $\tau$ in the rest frame of the $\tau$ and $P$ is the polarization, ranging from 0 to 1, of the sample of $\tau$s.  The parameter $\kappa$ is the spin analyzing power and ranges from $0$ to $\pm 1$.  A value of $\kappa=0$ indicates that there is no correlation between the spin of $\tau$ and the direction of the decay product chosen, while a value of $\kappa=\pm 1$ indicates the maximum correlation or anti-correlation.

Eq.~\eqref{eq:spin} is also the angular distribution of the decay product of a polarized $\tau$, which provides us with a method of computing it in Monte Carlo simulation. For our study, we generated a sample of polarized particles $\tau$ with the package \texttt{TauDecay} in \texttt{MadGraph 5}~\cite{Alwall:2011uj,Hagiwara:2012vz}.  The spin analyzing power is extracted from the distribution of Eq.~\eqref{eq:spin} by $\kappa=3\braket{\cos{\theta_d}}/P$. For the decay to $\rho$, we generate events of $\tau^-\rightarrow\nu_\tau\rho^-\rightarrow\nu_\tau\pi^0\pi^-$ and choose events with $\pi^0$ and $\pi^-$ whose invariant mass is $m_{\pi^0 \pi^-}=775.11\pm0.34~{\rm MeV}$ to reconstruct the $\rho^-$ invariant mass. 

\begin{table}
  \begin{center}
  \begin{tabular}{l|c|c|c}
  ~Decay channel~ &  &
  ~Branching fraction (\%)~ &
  ~Spin analyzing power~ \\
\hline
$\nu_\tau \pi^-$ & ~$\pi^-$~ & $10.8$ & $-1.00$\\
$\nu_\tau \rho^-$ & $\rho^-$ & $25.2$ & $0.45$\\
$\nu_\tau \pi^- \pi^+ \pi^-$ & $\pi^+$  & $9.3$ & $0.15$\\
$\nu_\tau \pi^- \pi^+ \pi^-$ & $\pi^-$  & $9.3$ & $0.04$ \\
$\nu_\tau \mu^- \bar{\nu}_\mu$ & $\mu^-$ & $17.4$ & $0.34$\\
$\nu_\tau e^- \bar{\nu}_e$ & $e^-$ & $17.8$ & $0.34$
\end{tabular}
\caption{The branching fractions and spin analyzing powers of the leading $\tau$ decay modes.}
\label{table:decays}
\end{center}
\end{table}

In Table~\ref{table:decays}, we list some of the potentially relevant decay channels of the $\tau$ together with the associated spin analyzing power.  Although due to the nature of the $\tau$'s interactions, all decays have an invisible $\nu_\tau$ in the final state, there are still enough constraints to reconstruct the four-vectors of each neutrino in some channels~\cite{Kuhn:1993ra,Fabbrichesi:2024wcd}.  At a glance, we see that the two most promising decay channels are $\tau^- \to \nu_\tau \pi^-$ which has a branching fraction of $11\%$ and a spin analyzing power of $1.00$ and $\tau^- \to \nu_\tau \rho^- \to \nu_\tau \pi^- \pi^0$ with a branching fraction of $25\%$ and a spin analyzing power of $0.45$.

In Sec.~\ref{sec:analysis}, we will perform a simple analysis using two methods.  The first is the decay method which utilizes spin correlations~\cite{Afik:2020onf}.  Consider the double differential cross section describing the angular distribution of one decay product of $\tau^-$ and one from $\tau^+$.
\begin{equation} \label{eq:diff2_xsec}
\frac{1}{\sigma}\frac{d^2\sigma}{d\cos\theta_{A,i} d\cos\theta_{B,j}}
= \frac{1}{4}\left( 1 
+ \kappa_A B^+_i \cos\theta_{A,i}
+ \kappa_B B^-_j \cos\theta_{B,j}
+ \kappa_A \kappa_B C_{ij} \cos\theta_{A,i} \cos\theta_{B,j}
\right). ~~~ 
\end{equation}
The angle $\theta_{A,i}$ is the angle of the $\tau^-$ decay product in the $\tau^-$ rest frame relative to the axis $i$ which is the spin quantization axis.  The angle $\theta_{B,j}$ is the angle of the $\tau^+$ decay product in the $\tau^+$ rest frame relative to the axis $j$ which is the spin quantization axis.  Each decay product has its associated spin analyzing power $\kappa_A$ or $\kappa_B$.  Finally, the coefficients $B^+_i$, $B^-_j$, and $C_{ij}$ are the same Fano coefficients as in Eq.~\eqref{eq:fano}.  

Integrating over various angles of Eq.~\eqref{eq:diff2_xsec} isolates the Fano coefficients:
\begin{subequations}
\label{eq:fano_by_decay}
\begin{align}
\label{eq:xsec_b+}
 \frac{1}{\sigma}\frac{d \sigma}{d\cos\theta_{A,i}}
& = \frac{1}{2} \left( 1 + \kappa_A B^+_i \cos\theta_{A,i} \right), \\
\label{eq:xsec_b-}
\frac{1}{\sigma}\frac{d \sigma}{d\cos\theta_{B,j}}
& = \frac{1}{2} \left( 1 + \kappa_B B^-_j \cos\theta_{B,j} \right), \\
\label{eq:xsec_cij}
\frac{1}{\sigma}\frac{d \sigma}{d\cos\theta_{A,i} \cos\theta_{B,j}}
& =- \frac{1}{2} \left( 1 + \kappa_A \kappa_B C_{ij} \cos\theta_{A,i} \cos\theta_{B,j} \right) \log | \cos\theta_{A,i} \cos\theta_{B,j} |.
\end{align}
\end{subequations}
Each Fano coefficient can be extracted from the measured distributions either by fitting, taking the asymmetry, or by taking the mean.  For instance, using the mean, the coefficients are extracted via:
\begin{align} 
B^+_i = \frac{3 \langle \cos\theta_{A,i} \rangle}{\kappa_A}, 
\qquad
B^-_j = \frac{3 \langle \cos\theta_{B,j} \rangle}{\kappa_B}, 
\qquad
C_{ij} = \frac{9 \langle \cos{\theta_{A,i}}\cos{\theta_{B,j}} \rangle}{\kappa_A \kappa_B}.
\end{align} 
Alternatively, the coefficients can be extracted using the asymmetry of the distribution since for each distribution in Eq.~\eqref{eq:fano_by_decay} the distribution is odd with respect to the differential variable.  The asymmetry $A(x)$ for a variable $x$ is given by
\begin{equation}
A(x)=\frac{N(x>0)-N(x<0)}{N(x>0)+N(x<0)},
\end{equation}
where $N(x>0)$ is the number of events with $x>0$ and $N(x<0)$ is the number of events with $x<0$.  With the asymmetry, the coefficients are
\begin{align} 
\label{eq:asymmetries}
B^+_i = \frac{2}{\kappa_A}A(\cos{\theta_{A,i}}),
\qquad
B^-_j = \frac{2}{\kappa_B}A(\cos{\theta_{B,j}}),
\qquad
C_{ij}=\frac{4}{\kappa_A\kappa_B} A(\cos{\theta_{A,i}}\cos{\theta_{B,j}}).
\end{align}

\section{$\tau^+ \tau^-$ at the BEPC}
\label{sec:BEPC}

\subsection{Center-of-Mass Energies}
\label{sec:com}

The Beijing Electron-Positron Collider II is an $e^+ e^-$ collider that has been operating since 2009.  Events are detected with the Beijing Spectrometer~\cite{BESIII:2022mxl,BESIII:2020nme}.  The experiment has produced outstanding results on $\tau$ physics and charm physics and has operated at center-of-mass energies ranging from $\sqrt{s} = 2.0-4.94~{\rm GeV}$ with an integrated luminosity of $35~{\rm fb}^{-1}$~\cite{BESIII:2022mxl}.

The BEPC-II is an excellent environment to measure final states with $\tau$s because the collider operates near the $\tau^+ \tau^-$ threshold, the backgrounds are very low, and the energy spread of the beam -- which is the dominant systematic uncertainty -- is very small, typically of the order $1-2~{\rm MeV}$.  BES-III has made a measurement of $\tau$ mass using $24~{\rm pb}^{-1}$ of data divided between four scan points near the pair production threshold with a statistical uncertainty of $0.1~{\rm MeV}$ and a systematic uncertainty of $0.1~{\rm MeV}$~\cite{BESIII:2014srs}.

The largest dataset of $\tau^+ \tau^-$ events comes from producing the $\psi(2S)$ resonance which subsequently decays into $\tau^+ \tau^-$.  The $\psi(2S)$ meson is a $c\bar{c}$ state with the quantum numbers $J^{PC} = 1^{--}$.  Its mass is $3686.097~{\rm MeV}$~\cite{KEDR:2003bik} and its branching fraction into $\tau^+ \tau^-$ is 0.31\%~\cite{ParticleDataGroup:2024cfk}.  The resulting $\tau$s have a velocity of $\beta=0.26$.  The number of $\psi(2S)$ events collected is $N_{\psi(2S)} = 2.7 \times 10^9$ leading to $N_{\psi(2S) \to \tau^+ \tau^-} = 3.5 \times 10^6$ events~\cite{BESIII:2024lks}.  The existing dataset is: 
\begin{itemize}
\vskip -0.1in 
  \item {\bf $\boldsymbol{\psi(2S)}$ dataset} with $\sqrt{s} = 3.686~{\rm GeV}$, $N_{\psi(2S) \to \tau^+ \tau^-} = 3.5 \times 10^6$, corresponding to $\beta = 0.26$.
\end{itemize}
The BEPC-II collider plans to upgrade their center-of-mass energy capabilities to $\sqrt{s} = 5.6~{\rm GeV}$ and collect a significant amount of data between the energies of $4.0~{\rm GeV}$ and $5.6~{\rm GeV}$. We consider two operational scenarios for our projections.
\begin{itemize}
  \item {\bf $\boldsymbol{5.6~{\rm GeV}}$ dataset} with $\sqrt{s} = 5.6~{\rm GeV}$, $\mathcal{L} = 20~{\rm fb}^{-1}$, corresponding to $\beta = 0.77$.

  \item {\bf $\boldsymbol{4.0 - 5.6~{\rm GeV}}$ dataset} with $\sqrt{s} = 4.0 - 5.6~{\rm GeV}$, with data taken at 20 different equally spaced center-of-mass energies with a cumulative integrated luminosity of $\mathcal{L} = 20~{\rm fb}^{-1}$, corresponding $\beta = 0.46 - 0.77$.
\end{itemize}

\subsection{Analysis}
\label{sec:analysis}

In this section, we present two simple analyses to measure entanglement and Bell nonlocality in $e^+ e^- \to \tau^+ \tau^-$ at BEPC-II.  In both analyses, the only selection cut we use is on the angle of the reconstructed $\tau$s.

The first analysis uses the {\bf decay method}.  In this method, we first select a particular decay of $\tau$.  We consider the $\tau^- \to \nu_\tau \pi^-$ decay because it has the maximal spin analyzing power, however, any reconstructable decay channel can be used or multiple channels can even be combined.  We then reconstruct the rest frame of each $\tau$.  The procedure involves with imposing the energy-momentum conservation between the initial $e^+e^-$ state and the final  $\tau^+\tau^-$ state, plus four more on-mass-shell conditions for $m_\tau$ and $m_\nu$, as outlined in Refs.~\cite{Kuhn:1993ra,Fabbrichesi:2024wcd}. Next, we boost to the rest frame of the $\tau^-$ and measure the angle of the $\pi^-$ with respect to the axes of the diagonal basis.

\begin{figure} 
  \centering
    \tikzset{every picture/.style={line width=0.75pt}} 
    \begin{tikzpicture}[x=0.75pt,y=0.75pt,yscale=-0.75,xscale=0.75]

        \draw  [draw opacity=0][line width=1.5]  (315.64,93.54) .. controls (326.78,98.22) and (336.17,106.25) .. (342.54,116.37) -- (293,147.5) -- cycle ; \draw  [line width=1.5]  (315.64,93.54) .. controls (326.78,98.22) and (336.17,106.25) .. (342.54,116.37) ;  
        \draw  [draw opacity=0][fill={rgb, 255:red, 0; green, 255; blue, 77 }  ,fill opacity=1 ] (274.26,149.57) -- (200.79,44.64) -- (198.94,45.93) -- (191.22,27.05) -- (206.32,40.77) -- (204.47,42.06) -- (277.95,146.99) -- cycle ;
        \draw  [draw opacity=0][fill={rgb, 255:red, 255; green, 0; blue, 0 }  ,fill opacity=1 ] (289.08,150.4) -- (394.01,76.93) -- (392.72,75.08) -- (411.6,67.36) -- (397.88,82.46) -- (396.59,80.61) -- (291.66,154.09) -- cycle ;
        \draw  [draw opacity=0] (304.26,113.15) .. controls (320.77,121.08) and (332.17,137.96) .. (332.17,157.5) .. controls (332.17,157.66) and (332.17,157.81) .. (332.16,157.97) -- (283,157.5) -- cycle ; \draw   (304.26,113.15) .. controls (320.77,121.08) and (332.17,137.96) .. (332.17,157.5) .. controls (332.17,157.66) and (332.17,157.81) .. (332.16,157.97) ;  
        \draw  [draw opacity=0][fill={rgb, 255:red, 0; green, 0; blue, 0 }  ,fill opacity=1 ] (281.67,166.83) -- (227.53,282.93) -- (229.57,283.88) -- (217.08,300.02) -- (221.41,280.08) -- (223.45,281.03) -- (277.59,164.93) -- cycle ;
        \draw  [draw opacity=0][fill={rgb, 255:red, 0; green, 0; blue, 0 }  ,fill opacity=1 ] (285,148.5) -- (339.14,32.4) -- (337.1,31.45) -- (349.59,15.32) -- (345.26,35.25) -- (343.22,34.3) -- (289.08,150.4) -- cycle ;
        \draw  [draw opacity=0][fill={rgb, 255:red, 131; green, 131; blue, 131 }  ,fill opacity=1 ] (125,155.25) -- (253.1,155.25) -- (253.1,153) -- (273,157.5) -- (253.1,162) -- (253.1,159.75) -- (125,159.75) -- cycle ;
        \draw  [draw opacity=0][fill={rgb, 255:red, 131; green, 131; blue, 131 }  ,fill opacity=1 ] (441,159.75) -- (312.9,159.75) -- (312.9,162) -- (293,157.5) -- (312.9,153) -- (312.9,155.25) -- (441,155.25) -- cycle ;
        \draw  [color={rgb, 255:red, 0; green, 148; blue, 255 }  ,draw opacity=1 ][line width=2.25]  (273,157.5) .. controls (273,151.98) and (277.48,147.5) .. (283,147.5) .. controls (288.52,147.5) and (293,151.98) .. (293,157.5) .. controls (293,163.02) and (288.52,167.5) .. (283,167.5) .. controls (277.48,167.5) and (273,163.02) .. (273,157.5) -- cycle ;
        \draw  [draw opacity=0][fill={rgb, 255:red, 0; green, 148; blue, 255 }  ,fill opacity=1 ][line width=2.25]  (277,157.5) .. controls (277,154.19) and (279.69,151.5) .. (283,151.5) .. controls (286.31,151.5) and (289,154.19) .. (289,157.5) .. controls (289,160.81) and (286.31,163.5) .. (283,163.5) .. controls (279.69,163.5) and (277,160.81) .. (277,157.5) -- cycle ;

        \draw (125,137) node [anchor=north west][inner sep=0.75pt]   [align=left] {{\fontfamily{ptm}\selectfont beam 1}};
        \draw (399,137) node [anchor=north west][inner sep=0.75pt]   [align=left] {{\fontfamily{ptm}\selectfont beam 2}};
        \draw (313.33,16.73) node [anchor=north west][inner sep=0.75pt]    {$\tau ^{-}$};
        \draw (237.33,273.4) node [anchor=north west][inner sep=0.75pt]    {$\tau ^{+}$};
        \draw (334.67,80.73) node [anchor=north west][inner sep=0.75pt]    {$\xi $};
        \draw (336,131.07) node [anchor=north west][inner sep=0.75pt]    {$\theta $};
        \draw (176.67,42.4) node [anchor=north west][inner sep=0.75pt]  [color={rgb, 255:red, 0; green, 255; blue, 77 }  ,opacity=1 ]  {$\overrightarrow{e_{1}}$};
        \draw (290.67,168.4) node [anchor=north west][inner sep=0.75pt]  [color={rgb, 255:red, 0; green, 148; blue, 255 }  ,opacity=1 ]  {$\overrightarrow{e_{2}}$};
        \draw (409.33,75.73) node [anchor=north west][inner sep=0.75pt]  [color={rgb, 255:red, 255; green, 0; blue, 0 }  ,opacity=1 ]  {$\overrightarrow{e_{3}}$};
    \end{tikzpicture}
  \caption{The diagonal basis of the $\tau^-\tau^+$ system near threshold.  It is related to the helicity basis by a rotation of $\xi$.}
  \label{fig:diag_basis}
\end{figure}
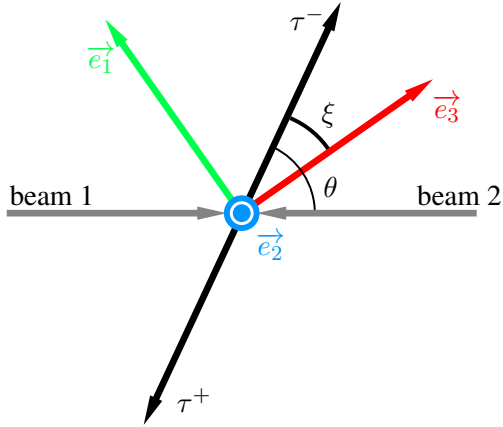

The simplest way to use this basis is to start from the helicity basis, defined with respect to the $\tau^-$ momentum direction $\vec k$, and then make an event-dependent rotation by $\xi$ in the scattering plane, as shown in Fig.~\ref{fig:diag_basis}.  In the low-energy limit, the angle is given by
\begin{equation}
\tan \xi = \sqrt{1-\beta^2}\ \tan \theta,
\end{equation}
where $\beta=\sqrt{1-4m_\tau^2/s}$ is the speed of the $\tau$ in the center-of-mass frame.  The same procedure is applied to $\tau^+$ to find the decay angle of $\pi^+$.

With the decay angles of $\pi^-$ and $\pi^-$ measured in the correct rest frames, we have measurements of the distributions of Eq.~\eqref{eq:fano_by_decay}.  The coefficients parameterizing the density matrix are extracted using Eq.~\eqref{eq:asymmetries}.  Having reconstructed the density matrix, we have characterized the $\tau^+ \tau^-$ quantum state and we can measure the concurrence with Eq.~\eqref{eq:concurrence_linear} and the Bell nonlocality with Eq.~\eqref{eq:bellnonlocality}.

The second analysis uses the {\bf kinematic method}.  With this method, we utilize the results of Sec.~\ref{sec:process} which show that the density matrix can be written as a function of the $\tau$ velocity $\beta$ and the scattering angle of the $\tau$ relative to the beam $\theta$.  In order to find $\beta$ and $\theta$, we need to reconstruct the four-momentum of $\tau^\pm$, but we do not need the decay angles.  We assume that we only use the $\tau^- \to \nu_\tau \pi^-$ decay channel, as discussed earlier in this section.  Although the kinematic method does not reconstruct spin proxies, a set of spin quantization axes is still chosen since this defines the parameterization, Eq.~\eqref{eq:fano}, of the density matrix.

\begin{figure}
  \centering
  \includegraphics[width=0.45\linewidth]{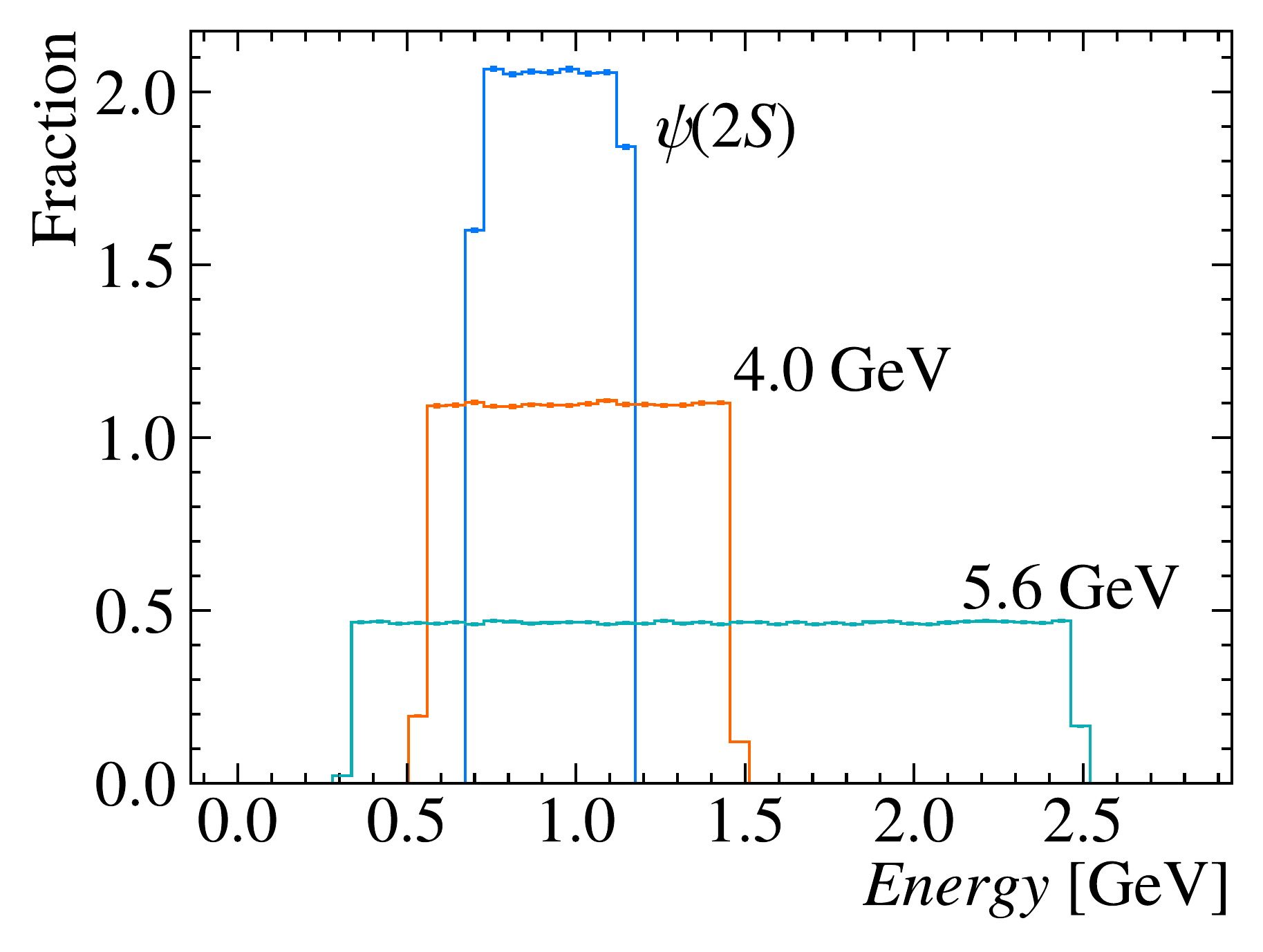}
  \qquad
  \includegraphics[width=0.45\linewidth]{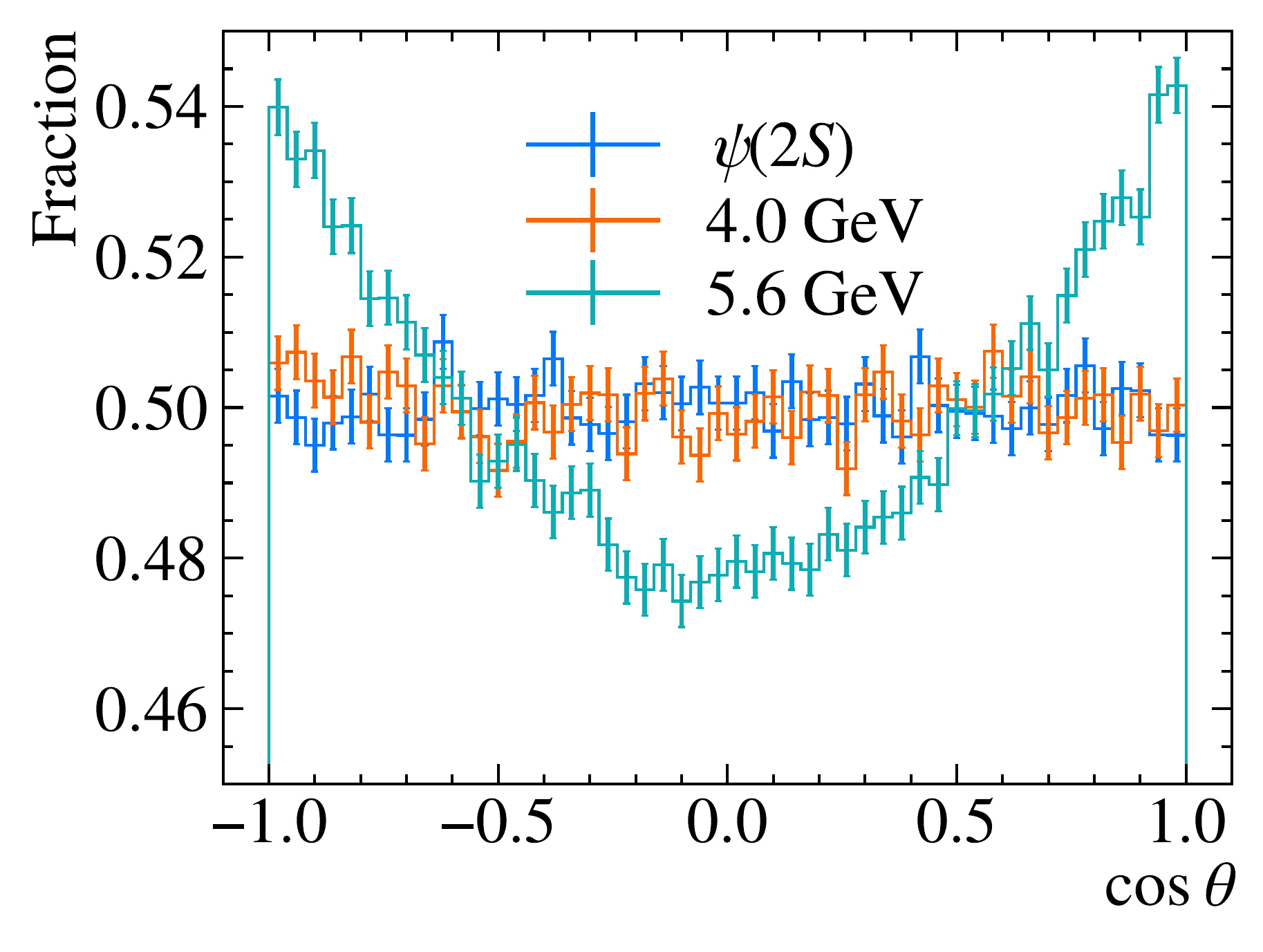}
  \caption{The energy spectrum (left) and angular distribution (right) of $\pi^\pm$ from the decays $\tau^- \to \nu_\tau \pi^-$ and $\tau^+ \to \bar{\nu}_\tau \pi^+$.   }
  \label{fig:pion_spectrum}
\end{figure}

Next, we discuss the practical aspects of the analysis.  For the decay method to sensibly extract the spin correlation coefficients from Eq.~\eqref{eq:fano_by_decay}, the distributions cannot be distorted.  Object selection, such as cuts on energy or transverse momentum introduces distortions.  Therefore, it is important to use the minimal possible object selection or to apply unfolding or an equivalent correction to remove the impact of such selections.  In the final state of $\nu_\tau \pi^- \bar{\nu}_\tau \pi^+$ that we consider, the only visible particles are the two pions.  Figure~\ref{fig:pion_spectrum} (left) shows the energy spectrum of pions while Fig.~\ref{fig:pion_spectrum} (right) shows the angular distribution, where $\theta$ is with respect to the beam.  Other BES-III studies have restricted charged tracks to $|\cos\theta|<0.93$~\cite{BESIII:2014srs,BESIII:2024lks}. 
We assume no object selection, but a reconstruction efficiency of $\epsilon_{\rm reco} = 0.9$, which is in line with previous BES-III work~\cite{BESIII:2024lks}.

\begin{figure}
  \centering
  \includegraphics[width=0.45\linewidth]{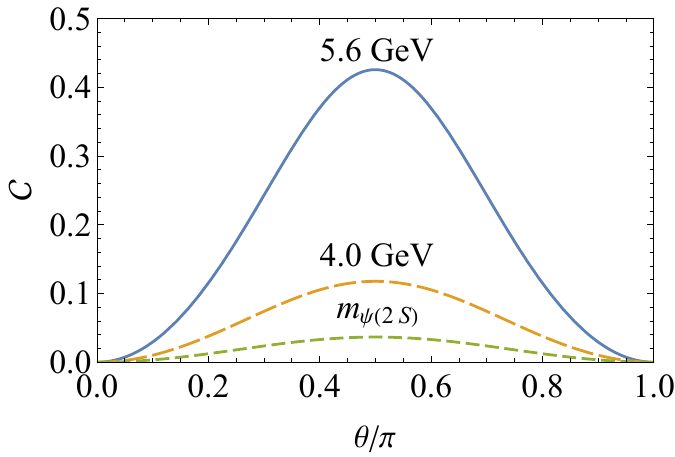}
  \qquad
  \includegraphics[width=0.45\linewidth]{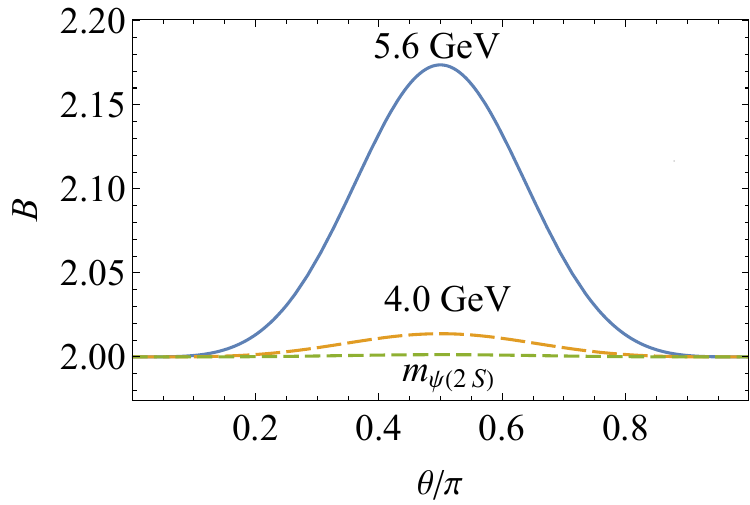}
  \caption{The concurrence (left) and Bell variable (right) as a function of scattering angle $\theta$ at the center-of-mass energies of $\sqrt{s} = m_{\psi(2S)} = 3.7~{\rm GeV}$, $\sqrt{s} = 4.0~{\rm GeV}$, and $\sqrt{s} = 5.6~{\rm GeV}$.}
  \label{fig:observable_angle}
\end{figure}

Cuts in the phase space of the scattering angle and collision energy do not distort distributions but simply define the quantum state that we are probing.  We wish to isolate a quantum state that exhibits as many quantum properties as possible.   Figure~\ref{fig:observable_angle} (left) shows the concurrence as a function of the scattering angle while Fig.~\ref{fig:observable_angle} (right) shows the Bell variable as a function of scattering angle.  The quantum behavior is the largest at $\theta = \pi/2$ which corresponds to $\tau$s that travel outwards perpendicular to the beam. Therefore, making a cut around the maximal angle, $\theta = \pi/2$, leads to a quantum state that is closest to a Bell state.  We thus select events that have reconstructed $\tau$s in an angular window of size $\Delta \theta$ around $\pi/2$
\begin{equation}
\label{eq:theta_cut}
\frac{\pi}{2} - \frac{\Delta \theta}{2} < \theta_\tau < \frac{\pi}{2} + \frac{\Delta \theta}{2} . 
\end{equation}
The value of $\Delta \theta$ can be optimized to achieve the optimal sensitivity based on the luminosity of the dataset.  A lower value of $\Delta \theta$ selects a state closer to a Bell state, but with lower statistics.  We note that $\Delta \theta$ may take a different value, and thus result in a different number of reconstructed events $N$, for the decay and kinematic methods to achieve their corresponding optimal sensitivity.

\begin{figure}
  \centering
  \includegraphics[width=0.42\linewidth]{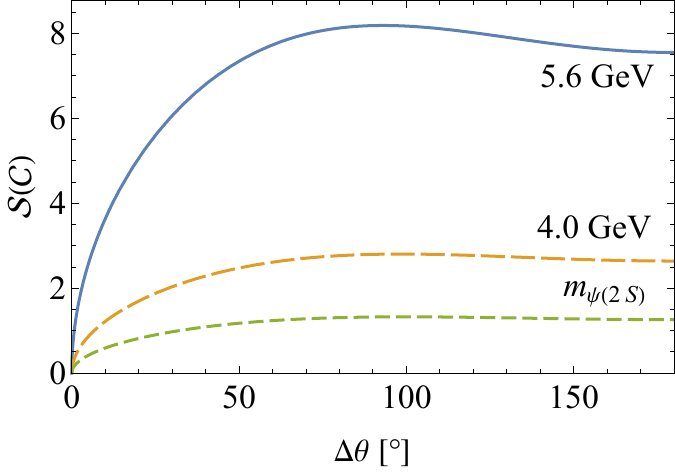}
  \qquad
  \includegraphics[width=0.45\linewidth]{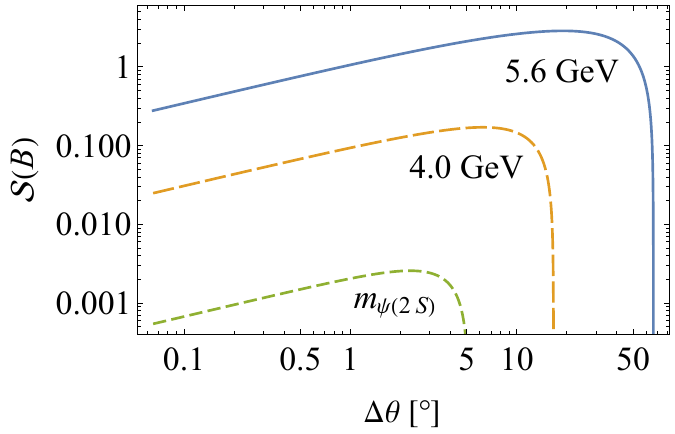}
  \caption{The significance of the concurrence (left) and Bell variable (right) with the decay method as a function of the size of the angular window $\Delta\theta$ at the center-of-mass energies of $\sqrt{s} = m_{\psi(2S)} = 3.7~{\rm GeV}$, $\sqrt{s} = 4.0~{\rm GeV}$, and $\sqrt{s} = 5.6~{\rm GeV}$ using the decay method with the $\nu\pi$ decay channel and the integrated luminosity $L=20\ \mathrm{fb^{-1}}$ for $\sqrt{s} = 4.0~{\rm GeV}$ and $\sqrt{s} = 5.6~{\rm GeV}$ and $N_{\psi(2S) \to \tau^+ \tau^-} =3.5\times10^6$ for $\sqrt{s} = m_{\psi(2S)}$. The systematic uncertainty $\Delta_{\rm sys}$ is not included. }
  \label{fig:theta_c}
\end{figure}

\subsection{Current Dataset}
\label{sec:current}

With the decay method, we measure the distributions of Eq.~\eqref{eq:fano_by_decay} using the cut $\Delta\theta$ described in Sect.~\ref{sec:analysis}.  The values of $B^+_i$, $B^-_j$, and $C_{ij}$ are extracted by the asymmetries in Eq.~\eqref{eq:asymmetries}.  The statistical uncertainty on Fano coefficients using the asymmetry is
\begin{equation}
\label{eq:stat_decay}
\Delta B^+_{i, \: {\rm stat}} = \frac{4}{\sqrt{N}},
\qquad\qquad
\Delta B^-_{j, \: {\rm stat}} = \frac{4}{\sqrt{N}},
\qquad\qquad
\Delta C_{ij, \: {\rm stat}} = \frac{4}{\sqrt{N}},
\end{equation}
where $N$ is the number of reconstructed events in the analysis and the numeric factor of 4 is from Eq.~\eqref{eq:asymmetries}.  

With the kinematic method, the statistical uncertainty is determined by the distribution of the quantity, with respect to the variables $\theta$ and $\beta$.  For example, $\Delta C_{11}$ is determined by the standard deviation of the $C_{11}$ entry of Eq.~\eqref{eq:spincorrelationmatrix}.  This uncertainty will also scale with $1/\sqrt{N}$ like Eq.~\eqref{eq:stat_decay}, but the pre-factor bounded to be less than 1~\cite{Cheng:2024rxi}, and in practice is often more than an order of magnitude smaller. 

These statistical uncertainties are propagated to uncertainties on $\mathcal{C}$ and $\mathcal{B}$. The final significance $\mathcal{S}$ is calculated as
\begin{equation}
\label{eq:significance}
\mathcal{S}(\mathcal{C}) = \frac{\mathcal{C}}{\sqrt{(\Delta \mathcal{C}_{\rm stat})^2+(\Delta \mathcal{C}_{\rm sys})^2}},
\qquad\qquad
\mathcal{S}(\mathcal{B}) = \frac{\mathcal{B} - 2}{\sqrt{(\Delta \mathcal{B}_{\rm stat})^2+(\Delta \mathcal{B}_{\rm sys})^2}}, 
\end{equation}
where $\Delta \mathcal{C}_{\rm sys}$ and $\Delta \mathcal{B}_{\rm sys}$ are the systematic uncertainties, which we generically refer to as $\Delta_{\rm sys}$. When it is relevant, we use the shorthand $\Delta \mathcal{C}_{\rm tot}=\sqrt{(\Delta \mathcal{C}_{\rm stat})^2+(\Delta \mathcal{C}_{\rm sys})^2}$ and $\Delta \mathcal{B}_{\rm tot}=\sqrt{(\Delta \mathcal{B}_{\rm stat})^2+(\Delta \mathcal{B}_{\rm sys})^2}$. 

The BEPC-II is a very clean experimental environment leading to systematic uncertainties that are controlled down to the sub-percent level for the $\tau^+ \tau^-$ cross section~\cite{BESIII:2024lks,BESIII:2017tvm,BESIII:2021cxx} and the $\tau$ mass~\cite{Mo:2016kfi,BESIII:2014srs}.  The energy scale is the leading uncertainty in a number of analyses and is known to $0.1~{\rm MeV}$~\cite{Mo:2016kfi}.  In other analyses, the overall systematic uncertainty is at the one to a few percent level~\cite{BESIII:2024xjf,BESIII:2024waz}.  We therefore choose to show our results for the following benchmark values of the systematic uncertainty as a percentage of the number of reconstructed events
\begin{equation}
\Delta_{\rm sys} = \{ 0.5\%,\quad 1\%,\quad 2\%,\quad 5\% \},
\end{equation}
which range from slightly optimistic to realistic to conservative.   In the following, we use $2\%$ as our default value to quote results.

In Fig.~\ref{fig:theta_c}, the significances $\mathcal{S}(\mathcal{C})$ and $\mathcal{S}(\mathcal{B})$ are shown with the decay method as a function of the angular window cut $\Delta \theta$ with a systematic uncertainty of $2\%$.  Operation at a higher center-of-mass energy is desirable to improve the observation because the $\tau^+ \tau^-$ state at higher energies is closer to a Bell state.

The results for the significances $\mathcal{S}(\mathcal{C})$ and $\mathcal{S}(\mathcal{B})$ in the current dataset, at the $\psi(2S)$ resonance, are shown in Table~\ref{table:results-psi2S}.  In the columns labeled $\Delta \theta$, we show the value of $\Delta \theta$ that would optimize the significance of the associated quantity.  However, in the significance calculation, we set a minimum angular window of $5^\circ$.  
Since the kinematic method measures elements of the spin correlation matrix event-by-event, the number of reconstructed events needed for a measurement can be very small resulting in a very small statistical uncertainty.  This is why the angular window optimization often leads to a very small preferred angular window size.
Furthermore, when the significance is above $5\sigma$, we show the predicted precision, corresponding to $1/\mathcal{S}$, achievable in the measurement.

\begin{table}
    \tabcolsep=0.2cm
    \centering
    \renewcommand\arraystretch{0.5}
        \begin{tabular}{|c|c||c|c|c|c||c|c|c|c|}
            \hline
            Method & $\Delta_{\rm sys}$   & $\mathcal{C}$     & $\Delta \mathcal{C}_{\rm tot}$ & $\mathcal{S}(\mathcal{C})$  & $\Delta\theta$ & $\mathcal{B}-2$     & $\Delta \mathcal{B}_{\rm tot}$ & $\mathcal{S}(\mathcal{B})$  & $\Delta\theta$\\
            \hline
            \multirow{5}[4]{*}[+0.5ex]{Decay}  & 0  & 0.029 & $ 0.020 $ & 1.42 & $100^\circ$ & $0.0011$ & 0.41 & $0.0026$ & $2.3^\circ$ \\
            \cline{2-10}
             & 0.5\% & $0.029$ & 0.021 & 1.42 & $100^\circ$ & 0.0011 & 0.41 & $0.0026$ & $2.3^\circ$ \\
            \cline{2-10}
             & 1\% & $0.029$ & 0.021 & 1.40 & $100^\circ$ & 0.0011 & 0.41 & $0.0026$ & $2.3^\circ$ \\
            \cline{2-10}
             & 2\% & $0.029$ & 0.022 & 1.33 & $100^\circ$ & 0.0011 & 0.42 & $0.0026$ & $2.3^\circ$ \\
            \cline{2-10}
             & 5\%  & 0.029 & 0.029 & 1.02 & $100^\circ$ & 0.0011 & 0.42 & $0.0025$ & $2.2^\circ$\\
            \hline\hline 
            \multirow{2}[4]{*}[-1.5ex]{Kinem.} & 0.5\% & $0.037$ & 0.0025 & $>5\sigma$, $7\%$ & 0 & 0.0013 & 0.010 & 0.13 & 0 \\
            \cline{2-10}
            & 1\% & $0.037$ & 0.0050 & $>5\sigma$, $14\%$ & 0 & 0.0013 & 0.020 & 0.067 & 0 \\
            \cline{2-10}
            & 2\% & $0.037$ & 0.010 & 3.70 & 0 & 0.0013 & 0.040 & 0.033 & 0 \\
            \cline{2-10}
             & 5\%  & 0.037 & 0.025 & 1.46 & 0 & 0.0013 & 0.10 & 0.013 & 0 \\
            \hline
        \end{tabular}
  \caption{The significance of observing entanglement and Bell nonlocality in the current $\psi(2S)$ dataset for the benchmark values of the systematic uncertanties $\Delta_{\rm sys}$, with the efficiency and cuts specified in Sec.~\ref{sec:analysis}.  When the optimal angular window is smaller than $5^\circ$, we use the non-optimal value of $5^\circ$.  When the significance is greater than $5\sigma$ we show the expected precision of the measurement $\mathcal{S}^{-1}$.} 
  \label{table:results-psi2S}
\end{table}

The first set of rows shows the results of the decay method for various systematic uncertainties.  The optimal angular window cut for concurrence is quite loose at $\Delta \theta = 100^\circ$ and leads to approximately $N \approx 281,000$ reconstructed events.  The concurrence is dominated by statistics, as evidenced by the weak dependence of $\Delta \mathcal{C}_{\rm tot}$ on $\Delta_{\rm sys}$.  Unfortunately, the signal of quantum entanglement is very weak and cannot be identified $\mathcal{C}>0$, using the decay method.  Bell nonlocality, $\mathcal{B}>2$, is even more difficult to establish.

The last set of rows in Table~\ref{table:results-psi2S} shows the results using the kinematic method for various systematic uncertainties.  The kinematic approach benefits from the less demanding reconstruction of the $\tau^+\tau^-$ events and from the more detailed form of kinematic distributions, Eq.~\eqref{eq:spincorrelationmatrix}.  The number of reconstructed events for the angular window cut of $5^\circ$, is $N\approx 1,590$.  Here, the systematic uncertainty plays a decisive role in the predicted sensitivity.  For a systematic uncertainty of $2\%$, a concurrence $\mathcal{C}>0$ would be observed at $3.7\sigma$.  To reach $5\sigma$ the systematic uncertainty needs to be controlled to $1\%$ or better.  Even with the kinematic method, the Bell nonlocality is still difficult to observe.

\subsection{Future Projections}
\label{sec:future}

\begin{table}
  \tabcolsep=0.2cm
  \centering
  \renewcommand\arraystretch{0.5}
  \begin{tabular}{|c|c||c|c|c|c||c|c|c|c|}
  \hline
             Method & $\Delta_{\rm sys}$   & $\mathcal{C}$     & $\Delta \mathcal{C}$ & $\mathcal{S}(\mathcal{C})$ & $\Delta\theta$ & $\mathcal{B}-2$     & $\Delta \mathcal{B}$ & $\mathcal{S}(\mathcal{B})$  & $\Delta\theta$ \\
            \hline
             \multirow{5}[4]{*}[+0.5ex]{Decay} & 0 & 0.33 & $0.039$ & $>5\sigma$, $12\%$ & $93^\circ$ & 0.14 & $0.031$ & 4.5 & $27^\circ$\\
            \cline{2-10}
            &0.5\% & 0.33 & 0.039 & $>5\sigma$, $12\%$ & $93^\circ$ & 0.14 & 0.039 & 4.3 & $25^\circ$\\
            \cline{2-10}
            &1\% & 0.33 & 0.039 & $>5\sigma$, $12\%$ & $93^\circ$ & 0.15 & 0.034 & 3.8 & $23^\circ$\\
            \cline{2-10}
            &2\% & 0.33 & 0.040 & $>5\sigma$, $12\%$ & $93^\circ$ & 0.16 & 0.054 & 2.9 & $19^\circ$\\
            \cline{2-10}
                & 5\%  & 0.33 & 0.044 & $>5\sigma$, $14\%$ & $93^\circ$& 0.17 & 0.11 & 1.5 & $12^\circ$ \\
              \hline\hline
              \multirow{4}[4]{*}[+1.0ex]{Kinem.} & 0.5\% & 0.43 & 0.0029 & $>5\sigma$, $0.68\%$ & $0^\circ$ & 0.17 & 0.0094 & $>5\sigma$, $5\%$ & $0^\circ$ \\
               \cline{2-10}
                &1\% & 0.43 & 0.0058 & $>5\sigma$, $1.4\%$ & $0^\circ$ & 0.17 & 0.019 & $>5\sigma$, $11\%$ & $0^\circ$ \\
                \cline{2-10}
                &2\% & 0.43 & 0.012 & $>5\sigma$, $2.7\%$ & $0^\circ$ & 0.17 & 0.037 & 4.7 & $0^\circ$ \\
                \cline{2-10}
                & 5\%  & 0.43 & 0.029 & $>5\sigma$, $6.7\%$ & $0^\circ$ & 0.17 & 0.094 & 1.8 & $0^\circ$ \\
            \hline
        \end{tabular}
  \caption{The significance of observing entanglement and Bell nonlocality in the future $5.6~{\rm GeV}$ dataset for the benchmark values of the systematic uncertanties $\Delta_{\rm sys}$, with the efficiency and cuts specified in Sec.~\ref{sec:analysis}.  When the optimal angular window is smaller than $10^\circ$, we use the non-optimal value of $10^\circ$.  When the significance is greater than $5\sigma$ we show the expected precision of the measurement $\mathcal{S}^{-1}$.} 
  \label{table:results-5.6}
\end{table}

\begin{table}
    \tabcolsep=0.2cm
    \centering
    \renewcommand\arraystretch{0.5}
        \begin{tabular}{|c|c||c|c|c|c||c|c|c|c|}
            \hline
             Method & $\Delta_{\rm sys}$  & $\mathcal{C}$     & $\Delta \mathcal{C}$ & $\mathcal{S}(\mathcal{C})$  & $\Delta\theta$ &  $\mathcal{B}-2$     & $\Delta \mathcal{B}$ & $\mathcal{S}(\mathcal{B})$  & $\Delta\theta$ \\
            \hline
             \multirow{5}[4]{*}[+0.5ex]{Decay} & 0  & 0.21 & $0.034$ & $>5\sigma$, $16\%$ & $96^\circ$ & $0.058$ & $0.026$ & 2.3 & $17^\circ$ \\
            \cline{2-10}
              & 0.5\% & 0.21 & 0.035 & $>5\sigma$, $16\%$ & $95^\circ$ & 0.060 & 0.028 & 2.1 & $16^\circ$ \\
              \cline{2-10}
                & 1\% & 0.21 & 0.035 & $>5\sigma$, $16\%$ & $95^\circ$ & 0.062 & 0.034 & 1.9 & $15^\circ$ \\
                \cline{2-10}
                & 2\% & 0.21 & 0.036 & $>5\sigma$, $17\%$ & $94^\circ$ & 0.066 & 0.050 & 1.3 & $12^\circ$ \\
                \cline{2-10}
                & 5\%  & 0.22 & 0.043 & $>5\sigma$, $20\%$ & $86^\circ$ & 0.070 & 0.10 & 0.67 & $7^\circ$ \\
            \hline\hline 
                 \multirow{3}[4]{*}[-2.0ex]{Kinem.} & 0  & 0.26 & 0.0010 & $>5\sigma$, $0.39\%$ & $45^\circ$ & 0.062 & 0.00018 & $>5\sigma, 0.29\%$ & $15^\circ$ \\
             \cline{2-10}
             & 0.5\% & 0.26 & 0.0029 & $>5\sigma$,  $1.1\%$ & $39^\circ$ & 0.073 & 0.0097 & $>5\sigma, 13\%$ & $1.2^\circ$ \\
             \cline{2-10}
            & 1\% & 0.26 & 0.0054 & $>5\sigma$, $2.0\%$ & $31^\circ$ & 0.073 & 0.019 & 3.8 & $0.77^\circ$ \\
            \cline{2-10}
            & 2\% & 0.27 & 0.011 & $>5\sigma$, $4.0\%$ & $21^\circ$ & 0.073 & 0.039 & 1.9 & $0.49^\circ$ \\
            \cline{2-10}
                  & 5\%  & 0.27 & 0.027 & $>5\sigma$, $9.9\%$ & $12^\circ$ &  0.073 & 0.097 & 0.75 & $0.26^\circ$ \\
            \hline
        \end{tabular}
  \caption{The significance of observing entanglement and Bell nonlocality in the future $4.0 - 5.6~{\rm GeV}$ dataset for the benchmark values of the systematic uncertanties $\Delta_{\rm sys}$, with the efficiency and cuts specified in Sec.~\ref{sec:analysis}.  When the optimal angular window is smaller than $10^\circ$, we use the non-optimal value of $10^\circ$.  When the significance is greater than $5\sigma$ we show the expected precision of the measurement $\mathcal{S}^{-1}$.} 
  \label{table:results-4.0-5.6}
\end{table}

In the immediate future, BEPC-II will continue its impressive physics mission and will upgrade its center-of-mass energy capabilities~\cite{BESIII:2022mxl}.  The BES-III experiment will collect a significant amount of data between the energies of $4.0~{\rm GeV}$ and $5.6~{\rm GeV}$.  We consider two operational scenarios as listed in Sec.~\ref{sec:com} for our projections. 

At higher center-of-mass energies, above the $\tau^+\tau^-$  threshold, the mixed state of $\tau^+\tau^-$ becomes more entangled, as discussed in Sec.~\ref{sec:qi}.  We show our results for the two energy-operation scenarios in Tables~\ref{table:results-5.6} and~\ref{table:results-4.0-5.6}, for both the decay and the kinematic methods, with different assumptions for the systematic uncertainties.  

First, at $\sqrt{s} = 5.6~{\rm GeV}$, we obtain $N\approx 1040-2300$ events for the decay method with angular window cuts of $\Delta \theta = 12^\circ - 27^\circ$, and $N \approx 860$ events for the kinematic method with angular window cuts of $\Delta \theta = 10^\circ$.  As shown in Table \ref{table:results-5.6}, the concurrence reaches $\mathcal{C} =  0.33 - 0.43$ corresponding to a substantially more entangled state than what is produced in the current $\psi(2S)$ dataset.  Consequently, the observation of entanglement is well above $5\sigma$.  Accordingly, we present the expected precision on a measurement of the concurrence.  We see that one would be able to reach $12\%$ accuracy from the decay method and about $3\%$ from the kinematic method with $\Delta_{\rm sys} = 2\%$. As for Bell nonlocality, the decay method would lead to an observation of $2.9\sigma$ and the kinematic method would lead to an observation of $4.7\sigma$.  If the systematics can be reduced to $\Delta_{\rm sys} < 2\%$ then $5\sigma$ is achievable.

For an alternative operational scenario in which the luminosity is distributed over many energy points, as shown in Table~\ref{table:results-4.0-5.6}, the concurrence reaches $\mathcal{C} = 0.21-0.27$, which is still quite sizable and allows for a $5\sigma$ observation with either method.  The expected precision reaches $17\%$ accuracy from the decay method and $4\%$ from the kinematic method with $\Delta_{\rm sys}=2\%$.  For the Bell nonlocality, a sensitivity of $2\sigma$ is achievable.  To reach an observation of $5\sigma$, the systematic uncertainty needs to be $0.5\%$ or better. 

\begin{figure}
  \centering
  \includegraphics[width=0.45\linewidth]{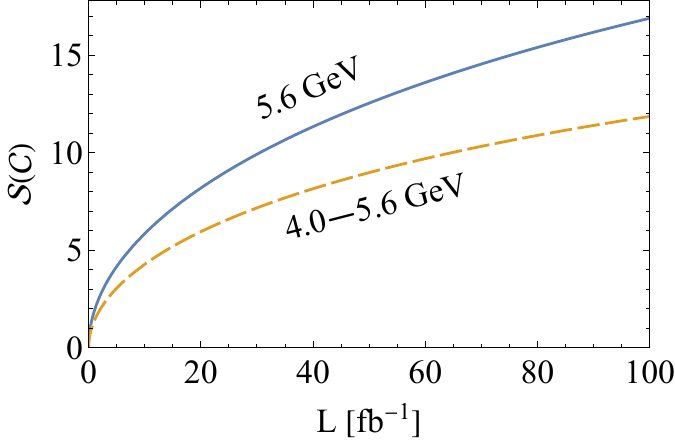}
  \qquad
  \includegraphics[width=0.45\linewidth]{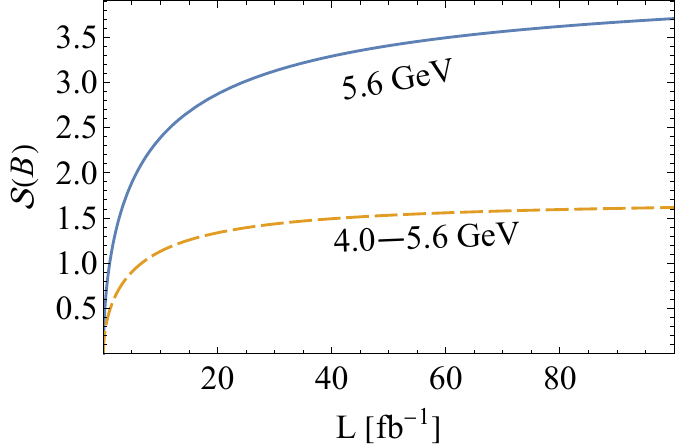}
  \caption{The optimal significance of observing entanglement (left) and Bell inequality violation (right) as a function of integrated luminosity at 5.6 GeV and $4.0-5.6$ GeV using the decay method in the $\nu\pi$ channel and $\Delta_{\rm sys}=2\%$.  The angular window cut is optimized at $L=20~{\rm fb}^{-1}$ and used for all luminosity values.}
  \label{fig:significance_vs_lumin}
\end{figure}

Finally, in Fig.~\ref{fig:significance_vs_lumin} we show the significance of concurrence (left) and Bell nonlocality (right) for the two energy operation scenarios as a function of luminosity using the decay method with $\Delta_{\rm sys}=2\%$.  The angular window cut is optimized for $20~{\rm fb}^{-1}$ and not re-optimized at different luminosity values.  The concurrence benefits from additional data while the Bell variable is nearly systematics dominated already with $20~{\rm fb}^{-1}$.  The measurements would be significantly improved using the kinematic method, as shown in Table~\ref{table:results-5.6} and Table~\ref{table:results-4.0-5.6}, for concurrence and for Bell nonlocality with low systematic uncertainty.  With the kinematic method, the precision is systematics dominated and consequently nearly independent of the integrated luminosity.  The concurrence would reach a precision of $3\%$ and $4\%$ in the operational scenarios of 5.6 GeV and $4.0-5.6$ GeV, respectively.  The significance of Bell nonlocality would reach $4.7\sigma$ and $1.9\sigma$, respectively, for the two operational scenarios.

\section{Summary and Conclusions}
\label{sec:summary}

There has been growing interest in studying the quantum tomography of quantum systems in high-energy collider experiments. In this work, we propose measuring quantum entanglement and Bell nonlocality in the $\tau^+\tau^-$ state near and above its kinematic threshold at BEPC-II.  We first laid out the procedure for quantum tomography of the $\tau^+\tau^-$ system produced in general $e^+e^-$ collisions and subsequent decays.  We introduced quantum quantities such as the concurrence and the Bell variable and discussed two approaches to construct them, namely the decay method and the kinematic method. 

The experimental program of BES-III is highly  acomplished and mature.  They have extensive experience and expertise in final states involving $\tau$ leptons and have already published a wide range of measurements, many of which have systematic uncertainties below $1\%$.  We found that in the existing dataset, consisting of $\psi(2S) \to \tau^+ \tau^-$, entanglement is observable when systematic uncertainties are at or below $1\%$.

In the upcoming run between 4.0 GeV and 5.6 GeV, we presented two energy-operation scenarios.  While the higher-energy operation would be more beneficial for the observation of quantum effects, both cases will significantly improve the quantum tomography of the $\tau^+ \tau^-$ state.  Entanglement is not only observable in this situation, but can be measured with a precision of $4\%$ or better.

Bell nonlocality is also potentially observable but requires a bit more control over systematic uncertainties.  If the full dataset is collected at 5.6 GeV, then the control below $2\%$ is sufficient for a $5\sigma$ observation and if the data are distributed between 4.0 GeV and 5.6 GeV, then the control down to $0.5\%$ is needed.

BES-III already has an impressive physics program covering a wide array of topics.  In this work, we have shown that they can extend their program to quantum information with a few straightforward measurements.  The entry point into quantum information typically starts with measuring entanglement and Bell nonlocality.  Since quantum tomography is possible, many other quantum quantities can also be explored, such as quantum discord, steering, or negative conditional entropy.  These measurements would be a challenge at BEPC-II, since the $\tau^+ \tau^-$ quantum state has a substantial separable component when operating at an energy not far above the threshold.

\section*{Acknowledgments}

The authors thank Roy Briere, Kun Cheng, and Jiayin Gu for helpful discussions.  This work was supported in part by the U.S.~Department of Energy under grant No.~DE-SC0007914 and in part by Pitt PACC.  ML is also supported by the National Science Foundation under grant No.~PHY-2112829.

\appendix
\section{Phenomenological Description of $\psi(2S)$}
\label{sec:psi2S}

\begin{figure}
    \centering
    \includegraphics[width=0.45\linewidth]{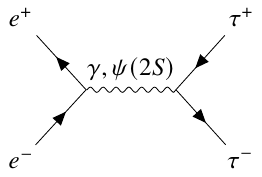}
    \caption{The Feynman diagram of the process $e^+ e^- \to \tau^+ \tau^-$ near $\psi(2S)$ resonance.}
    \label{fig:feynman2}
\end{figure}

The $\psi(2S)$ is a vector meson resonance of $c\bar{c}$ with the same $J^{PC}$ quantum numbers as the photon. We therefore include both mediators to produce the $\tau^+ \tau^-$ state as shown in Fig.~\ref{fig:feynman2}.  In addition to the mass and total width, as well-measured parameters $m_V = 3686.097$ MeV and $\Gamma_V = 293$ keV~\cite{ParticleDataGroup:2024cfk}, we introduce its vector couplings to leptons
\begin{equation}
\mathcal{L}_{\rm int} \supset  g_{V,\tau} V^\mu \bar{\tau} \gamma_\mu \tau 
+ g_{V,\mu} V^\mu \bar{\mu} \gamma_\mu \mu
+ g_{V,e} V^\mu \bar{e} \gamma_\mu e.
\end{equation}
The total cross section near the $\psi(2S)$ resonance is thus expressed as 
\begin{equation}
    \sigma(e^+ e^- \to \tau^+ \tau^-)=\frac{s}{12 \pi}\left|\frac{g_{V,\tau} g_{V,e}}{\left(s-m_V^2\right)+i m_V \Gamma_V}+\frac{e^2}{s}\right|^2 \sqrt{1-\frac{4 m_\tau^2}{s}}\left(1+\frac{2 m_\tau^2}{s}\right) . 
    \label{eq:cross_section}
\end{equation}
The peak cross section of this process near the $\psi(2S)$ resonance measured in BES-III is $52~\mathrm{nb}$~\cite{Achasov:2019rdp} which lets us obtain the product of couplings $g_{V,\tau} g_{V,e} = 3.1\times10^{-5}$.  Following the vector meson dominance model~\cite{Sakurai:1960ju}, assuming that $g_{V,\tau} = g_{V,e}$ leads to calculated branching fraction of ${\rm BF}(\psi(2S) \to e^+e^-) =  1\%$ and ${\rm BF}(\psi(2S) \to \tau^+\tau^-) =  0.4\%$ which are consistent with the measured values at the 20\% level~\cite{ParticleDataGroup:2024cfk}.  This parametrization can be used to describe $\tau^+ \tau^-$ production, even off the $\psi(2S)$ resonance.

\section{Fictitious States and Results in the Beam Basis}
\label{sec:fictitious}

In colliders, when an event-dependent basis is used,  instead of reconstructing the Fano coefficients from Eq.~\eqref{eq:fano}, the averaged Fano coefficients are reconstructed~\cite{Afik:2022kwm}.  This is due to the angular integration -- including events in the analysis that have different values of $\phi$ -- which is nearly always the case in collider experiments.  The use of averaged Fano coefficients also causes the results to depend on the spin quantization basis used.  A reconstructed quantum state that is basis-dependent is called a fictitious state~\cite{Afik:2022kwm,Cheng:2023qmz,Cheng:2024btk}.

The helicity basis, used in our main analysis, is an event-dependent basis because the quantization axes are set by the $\tau$ momentum direction $\hat{k}$, which is different for each event.  Fortunately, it has been shown that if concurrence or Bell nonlocality is non-zero in a fictitious state, then there exists a substate that also has non-zero concurrence or Bell nonlocality~\cite{Afik:2022kwm,Cheng:2023qmz,Cheng:2024btk}.  With appropriate cuts, a corresponding genuine quantum state can be shown to be entangled or Bell nonlocal.

In this appendix, we provide results for the future 5.6 GeV dataset in the beam basis which is a fixed basis for lepton colliders.  When this basis is used, the reconstructed density matrix represents a genuine quantum state. The resulting significance is inferior to that using the helicity basis.  

For a single point in phase space $(\theta,\phi)$ the Fano coefficients, $C_{ij}^{\rm beam}$ and $B_i^{{\rm beam}, \pm}$, in the beam basis can be obtained from the Fano coefficients in the helicity basis, $C_{ij}^{\rm helicity}$ and $B_i^{{\rm helicity},\pm}$, by event-dependent rotation $R(\theta,\phi)$.
\begin{equation}
C_{ij}^{\rm beam }=
R^T(\theta,\phi)_{ik}C^{\rm helicity}_{kl} R(\theta,\phi)_{lj},
\qquad 
B_i^{{\rm fixed},\pm}=
R^T(\theta,\phi)_{ij}B^{{\rm helicity},\pm}_{j}.
\end{equation}
In practice, we typically study the states averaged over the azimuthal angle $\phi$.  In the helicity basis this averaging leads to a fictitious state, but in the beam basis this averaging retains a quantum state.  In the beam basis averaged over $\phi$, the Fano coefficients take the form
\begin{equation}
    C_{ij}=\left (\begin{array}{ccc}
        C_1 & 0 & 0 \\
        0 & C_1 & 0 \\
        0 & 0 & C_3
    \end{array}\right )_{ij},
    \qquad
    B^+_i=B^-_i=\left (\begin{array}{c}
        0 \\
        0  \\
        B_3 
    \end{array}\right )_{i}.
\end{equation}
The corresponding value of the concurrence is
\begin{equation}
\mathcal{C}=\left\{
\begin{array}{ll}
0, 
& \qquad\text{if } C_3>C_1+B_3^2/(C_1+1), \\
\frac{1}{4}(1+2C_1-C_3)-\frac{1}{2}\sqrt{(1+C_3)^2-4B_3^2}, 
& \qquad\text{if } C_3<C_1+B_3^2/(C_1+1).
\end{array}
\right. 
\end{equation}
The Bell variable is 
\begin{equation}
\mathcal{B}=\left\{
\begin{array}{ll}
2\sqrt{C_1^2+C_3^2},
& \qquad\text{if } |C_3|>|C_1|, \\
2\sqrt{2}|C_1|, 
& \qquad\text{if } |C_3|<|C_1|.
\end{array}
\right.
\end{equation}
These can be converted from the helicity basis, for example from Eq.~\eqref{eq:fano_helicity}, according to
\begin{subequations}
\begin{align}
 C_1&=\frac{1}{2}(C_{11}^{\rm helicity}\cos^2{\theta}+C_{33}^{\rm helicity}\sin^2{\theta}-2C_{13}^{\rm helicity}\cos{\theta}\sin{\theta}+C_{22}^{\rm helicity}), \\
C_3&=C_{11}^{\rm helicity}\sin^2{\theta}+C_{33}^{\rm helicity}\cos^2{\theta}+2C_{13}^{\rm helicity}\cos{\theta}\sin{\theta}, \\
B_3
& =B_1^{\rm helicity}\sin{\theta}+B_3^{\rm helicity}\cos{\theta}.
\end{align}
\end{subequations}
The beam basis is described by $\left \{C_1,C_3,B_3 \right \}$ and the helicity basis is described by \\$\left \{C_{11}^{\rm helicity},C_{13}^{\rm helicity},C_{22}^{\rm helicity},C_{33}^{\rm helicity},B^{\rm helicity}_1,B^{\rm helicity}_3 \right \}$.

\begin{figure}
  \centering
  \includegraphics[width=0.45\linewidth]{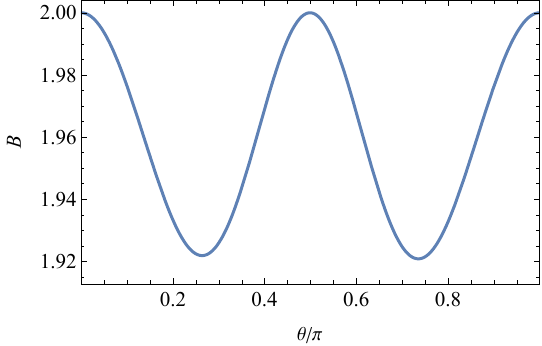}
  \caption{The Bell variable in the beam basis as a function of scattering angle $\theta$ at the center-of-mass energy of $\sqrt{s}=5.6\ \mathrm{GeV}$.  A value of $\mathcal{B}>2$ indicates Bell nonlocality.}
  \label{fig:fixed_Bell}
\end{figure}

For energies within reach of BEPC-II, $C_3>C_1+ B_3^2/(C_1+1)$ for any scattering angle $\theta$, which means that the concurrence is 0, and the $\tau^+ \tau^-$ state is separable.  The situation is the same for Bell nonlocality.  Figure~\ref{fig:fixed_Bell} shows the Bell variable, in the beam basis, as a function of the scattering angle $\theta$ at the center-of-mass energy of $\sqrt{s}=5.6~\mathrm{GeV}$.  For all values of $\theta$ the $\tau^+ \tau^-$ state is Bell local.

Since entanglement and Bell nonlocality is observable in the helicity basis, in Sec.~\ref{sec:BEPC}, there exists a quantum state which is entangled and Bell nonlocal~\cite{Cheng:2023qmz,Cheng:2024btk}.  While in the beam basis, choosing the phase space which integrates over the full azimuthal region leads to a separable state, restricting the phase space to a smaller region in the azimuthal angle $\phi_\tau$ can enhance some of the quantum substates that are entangled and Bell nonlocal.

In particular, we perform an angular window cut in the azimuthal direction of the $\tau$
\begin{equation}
\label{eq:phi_cut}
0 < \phi_\tau < \frac{\pi}{4}.
\end{equation}
We show the results at 5.6 GeV using this azimuthal cut and additionally optimizing the $\Delta \theta$, as in Eq.~\eqref{eq:theta_cut}, in Table~\ref{table:results-fixed}.  Using the decay method the results are quite a bit worse than the results in the helicity basis in Table~\ref{table:results-5.6}.  With the kinematic method, on the other hand, the results are still worse than in the helicity basis, but only slightly.

\begin{table}
  \tabcolsep=0.2cm
  \centering
  \renewcommand\arraystretch{0.5}
  \begin{tabular}{|c|c||c|c|c|c||c|c|c|c|}
  \hline
             Method & $\Delta_{\rm sys}$   & $\mathcal{C}$     & $\Delta \mathcal{C}$ & $\mathcal{S}(\mathcal{C})$ & $\Delta\theta$ & $\mathcal{B}-2$     & $\Delta \mathcal{B}$ & $\mathcal{S}(\mathcal{B})$  & $\Delta\theta$ \\
            \hline
             \multirow{5}[4]{*}[+0.5ex]{Decay} & 0 & 0.20 & 0.016 & $>5\sigma$, $7.7\%$ & $98^\circ$  & 0.11 & 0.089 & 1.22 & $50^\circ$ \\
            \cline{2-10}
            &0.5\% & 0.20 & 0.016 & $>5\sigma$, $7.8\%$ & $97^\circ$  & 0.11 & 0.090 & 1.21 & $49^\circ$\\
            \cline{2-10}
            &1\% & 0.21 & 0.017 & $>5\sigma$, $8.1\%$ & $92^\circ$  & 0.11 & 0.093 & 1.19 & $48^\circ$\\
            \cline{2-10}
            &2\% & 0.22 & 0.020 & $>5\sigma$, $9.1\%$ & $80^\circ$  & 0.11 & 0.10 & 1.11 & $45^\circ$\\
            \cline{2-10}
                & 5\% & 0.24 & 0.033 & $>5\sigma$, $14\%$ & $55^\circ$  & 0.12 & 0.15 & 0.84 & $35^\circ$ \\
              \hline\hline
              \multirow{4}[4]{*}[+1.0ex]{Kinem.} & 0.5\% & 0.27 & 0.0027 & $>5\sigma$, $0.99\%$ & 0  & 0.14 & 0.0097 & $>5\sigma$, $6.8\%$ & 0 \\
               \cline{2-10}
                &1\% & 0.27 & 0.0054 & $>5\sigma$, $2.0\%$ & 0  & 0.14 & 0.019 & $>5\sigma$, $14\%$ & 0 \\
                \cline{2-10}
                &2\% & 0.27 & 0.011 & $>5\sigma$, $4.0\%$ & 0  & 0.14 & 0.039 & 3.66 & 0 \\
                \cline{2-10}
                & 5\% & 0.27 & 0.027 & $>5\sigma$, $9.9\%$ & 0  & 0.14 & 0.097  & 1.46 & 0 \\
            \hline
        \end{tabular}
  \caption{The significance of observing entanglement and Bell nonlocality using the fixed beam basis with azimuthal cut  $0 < \phi_\tau  < \pi/4$ in the future $5.6~{\rm GeV}$ dataset for the benchmark values of the systematic uncertanties $\Delta_{\rm sys}$, with the efficiency and cuts specified in Sec.~\ref{sec:analysis}.  When the optimal angular window $\Delta\theta$ is smaller than $10^\circ$, we use the non-optimal value of $10^\circ$.  When the significance is greater than $5\sigma$ we show the expected precision of the measurement.} 
  \label{table:results-fixed}
\end{table}

\bibliographystyle{apsrev4-1}
\bibliography{refs}
\end{document}